\documentclass[aps,prr,twocolumn,showpacs,preprintnumbers,amsmath,amssymb,
floatfix,superscriptaddress]{revtex4-1}

\usepackage{graphics}
\usepackage{graphicx}
\usepackage{amsmath}
\usepackage{setspace}
\usepackage{braket}
\usepackage{epstopdf}
\usepackage{comment}
\usepackage{hyperref}
\usepackage{bm}
\usepackage{xfrac}

\usepackage{mathrsfs}

\hypersetup{%
   pdfpagemode=None, 
   pdfstartpage=1,
   pdfmenubar=true,
   pdftoolbar=true,
   colorlinks = true,
   linkcolor=blue,
   citecolor=blue,
   urlcolor=blue,
   bookmarksopen=false
 }

\newcommand{\affA}{Van der Waals-Zeeman Institute, Institute of Physics, University of Amsterdam, 1098 XH Amsterdam, Netherlands}
\newcommand{\affB}{Institute for Molecules and Materials, Radboud University, Heyendaalseweg 135, 6525 AJ Nijmegen, Netherlands}
\newcommand{\affC}{Fritz-Haber-Institut der Max-Planck-Gesellschaft, Faradayweg 4-6, 14195 Berlin, Germany }
\begin{document}

\title{Controlling the nature of a charged impurity in a bath of Feshbach dimers}


\author{Henrik~Hirzler}\affiliation{\affA}
\author{Eleanor~Trimby}\affiliation{\affA}
\author{Rianne\,S.~Lous}\affiliation{\affA}
\author{Gerrit\,C.~Groenenboom}\affiliation{\affB}
\author{Rene~Gerritsma}\affiliation{\affA}
\author{Jes\'{u}s~P\'{e}rez-R\'{i}os}\affiliation{\affB}\affiliation{\affC}

\date{\today}

\begin{abstract}
We theoretically study the dynamics of a trapped ion that is immersed in an ultracold gas of weakly bound atomic dimers created by a Feshbach resonance. Using quasi-classical simulations, we find a crossover from dimer dissociation to molecular ion formation depending on the binding energy of the dimers. The location of the crossover strongly depends on the collision energy and the time-dependent fields of the Paul trap. Deeply bound dimers lead to fast molecular ion formation, with rates approaching the Langevin collision rate $\Gamma'_\text{L}\approx4.8\times10^{-9}\,$cm$^3$s$^{-1}$. The kinetic energies of the created molecular ions have a median below $1\,$mK, 
such that they will stay confined in the ion trap. We conclude that interacting ions and Feshbach molecules may provide a novel approach towards the creation of ultracold molecular ions with applications in precision spectroscopy and quantum chemistry.
\end{abstract}

\maketitle

\section{Introduction}
Recently, trapped ions have been combined with ultracold atomic gases~\cite{Zipkes:2010,Schmid:2010,Harter:2013,Meir:2016,Cote:2016,Haze:2018,Tomza:2019,Mohammadi:2020}. These systems are of particular interest to study charged impurity physics in a quantum bath. The well-controlled ionic impurities may be used to probe properties of the atomic bath, or to study the decoherence of internal states and motion while interacting with a quantum environment~\cite{Gorden:1979,Daley:2004,Kollath:2007,Ratschbacher:2013,Meir:2016,Kleinbach:2018,Feldker:2020,Schmidt:2020}. Notably, the charge-dipole interactions are longer-ranged than those found in neutral systems~\cite{Casteels:2011,Tomza:2019}. 
This could lead to larger polaronic effects~\cite{Astrakharchik:2020} and it has been suggested that many atoms can become weakly bound to a single ion~\cite{Cote:2002}. The system is experimentally attractive as both the motion and internal states of individual trapped ions can be accurately controlled and measured \cite{Monroe:1995a,Leibfried:2003}. Furthermore, the interactions within the atomic bath can be tuned with Feshbach resonances and these even allow for transforming the bath into a gas of molecules \cite{Chin:2010}. However, the long-range interactions tend to translate into a higher reactivity, as has been shown for instance for Rydberg impurities in an atomic gas~\cite{Schlagmueller:2016}. Therefore, understanding the chemistry of the species involved is fundamental to develop models for charged impurities in ultracold gases. 

In this work, we present a theoretical study of a single ion impurity in a bath of ultracold diatomic molecules whose binding energy can be controlled with a Feshbach resonance. 
We show that a crossover exists in the system, depending on the molecular binding energy $E_\text{b}$ and the ion-molecule collision energy $E_\text{col}$. We find that a charged impurity reacts with a molecule of the bath leading, as the main reaction channel, to the formation of molecular ions, which can be viewed as a charged-molecular impurity. However, as soon as $E_\text{col} \gg E_\text{b}$, the impurity predominantly induces the dissociation of the dimer, and the molecular ion creation rate drops significantly. In other words, by tuning the binding energy it is possible to control the nature of a charged impurity.  Our results open a new avenue towards the creation of ultracold molecular ions with applications in quantum chemistry and precision spectroscopy \cite{MurPetit:2012,Wolf:2016,Chou:2017, Sinhal:2020,Mohammadi:2020}.

\begin{figure*}
	\centering
	\includegraphics[width=0.78\textwidth]{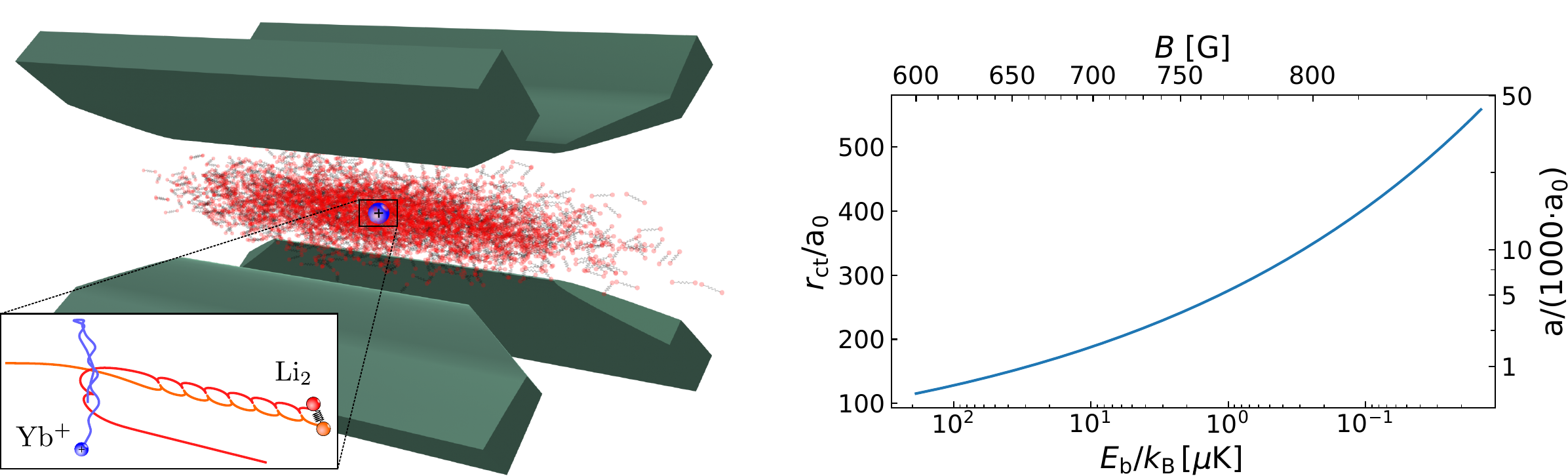}
 	\caption{Left: Schematics of the atom-ion system investigated. A single ion (blue) trapped in the radio-frequency blades of a Paul trap (gray) combined with a cloud of ultracold molecules (red). Inset:~Sketch of a single simulation run. A Yb$^+$ ion (blue) oscillating in the trapping field of the Paul trap collides with a Li$_2$ dimer (red, orange), whereby the Li$_2$ is dissociated. Right: Classical turning point $r_\mathrm{ct}$, binding energy $E_\text{b}$ and scattering length $a$ as a function of the magnetic field strength $B$ for Li$_2$ dimers created close to the Feshbach resonance.
 	 }
 	\label{fig:mol_size}
\end{figure*}

As a prime example, we study the $^6$Li$_2$-Yb$^+$ system inside the radio frequency electric fields of a Paul  trap as sketched in Fig.~\ref{fig:mol_size}. The large mass ratio is appealing to study chemical reactions experimentally, as Yb$^+$ and LiYb$^+$ can be confined simultaneously despite the Paul trap acting as a mass filter. Furthermore, the mass ratio mitigates adverse heating effects from the Paul trap \cite{Cetina:2012,Fuerst:2018}, which allowed to reach ultracold atom-ion collision energies on the order of $10\,\mu$K~\cite{Feldker:2020}. For the reported collision energies the full crossover regime is within experimental reach. 
The $^6$Li atoms feature a broad Feshbach resonance around $832\,$G  between the two lowest energy spin states~\cite{Zuern:2013}. On the repulsive side of this resonance, long-lived Li$_2$ dimers are produced by three-body recombination once the atoms are sufficiently cold \cite{jochim:2003}. The binding energy of these weakly bound dimers lies in the $\mu$K range and can be straightforwardly tuned using an external magnetic field.

\section{Theory}
We simulate the dynamics of the colliding ions and molecules using the quasi-classical trajectory (QCT) method. This  approach has been used to treat scattering problems in the chemical physics community since the pioneering work of Karplus et al.~\cite{Karplus} and it has  recently been applied to the study of cold chemical reactions between molecular ions and neutrals~\cite{Perez:2019}. The QCT method calculates the trajectories classically but the initial conditions of the colliding partners are selected according to the quantum state of the reactants through the celebrated Wentzel, Kramers, and Brillouin (WKB) or semi-classical approximation~\cite{Truhlarbook,Levine}. QCT is applied in scattering problems where many partial waves contribute~\cite{Perez:2019} or when the problem is too complex for a full quantum treatment. The latter is the case when we consider the electric fields of a Paul trap, which have a massive impact on atom-ion scattering \cite{Weckesser:2015, Pinkas:2020, Fuerst:2018}, but severely complicate calculations due to its asymmetry and explicit time dependence. 

The potential of a Paul trap is given by:
\begin{equation}
	V(\vec{r},t) = \frac{U_\mathrm{dc}}{2} \sum_{j=1}^3\alpha_j r_{\text{i}_j}^2 + \frac{U_\mathrm{rf}}{2} \cos\left(\Omega t \right) \sum_{j=1}^3 \alpha_j^\prime r_{\text{i}_j}^2,
\end{equation}
with $\vec{r}_{\text{i}}=\vec{0}$ the center of the trap,  $U_\mathrm{dc}$ and $U_\mathrm{rf}$ curvatures of static and radio-frequency fields respectively and geometry factors $\alpha_j^{(\prime)}$. For the linear Paul trap considered here, $-2\alpha_1=-2\alpha_2=\alpha_3=1$ and $\alpha_1^{(\prime)}=-\alpha_2^{(\prime)}=1$, $\alpha_3^{(\prime)}=0$. The motion of the ion in the transverse directions can be descibed by a slow (secular) motion with a frequency $\omega_{\perp}\approx \Omega q/\sqrt{8}$ and $q=2eU_\mathrm{rf}/(m_{\text{Yb}^+}\Omega^2)$ superimposed on a fast micromotion with frequency $\Omega$~\cite{Leibfried:2003}. The motion along the axial $z$-direction is purely harmonic with frequency $\omega_3=\sqrt{eU_\mathrm{dc}/m_{\text{Yb}^+}}$.

The atom-ion potential consists of a characteristic long range term {$C_4^\text{ai}/r_\text{ai}^4$ which is a consequence of the charge-induced dipole interaction. ${ C_4^\text{ai}}$ is the attractive interaction coefficient } and $r_\text{ai}$ the atom-ion distance. 
Two atom-ion collision types can be distinguished: large impact parameters $b$ lead to elastic scattering (glancing collisions), whereas for $b<b_\mathrm{c}=(2C_4^\text{ai}/E_\mathrm{col})^{1/4}$ spiraling Langevin collisions occur in which large momentum and energy transfer is possible \cite{Langevin:1905}. The Langevin collision rate $\Gamma_\mathrm{L}=2\pi\rho_\mathrm{a}\sqrt{C_4^\text{ai}/\mu_\mathrm{ai}}$ is independent of the collision energy. Here, $\rho_\mathrm{a}$ is the cloud density and $\mu_\mathrm{ai}$ is the atom-ion reduced mass. 
We model the atom-ion potential with
\begin{equation}
	V_\mathrm{ai}(r_\mathrm{ai}) = -\frac{C_4^{\mathrm{ai}}}{2 r_\mathrm{ai}^4} +\frac{C_6^{\mathrm{ai}}}{r_\mathrm{ai}^6}, \quad  r_\mathrm{ai} = \left|\vec{r}_\mathrm{a}-\vec{r}_\mathrm{i}\right|,
\end{equation}
where $\vec{r}_\mathrm{a}$ and $\vec{r}_\mathrm{i}$ are the atom and ion position respectively and $C_6^{\mathrm{ai}}$ is the repulsion coefficient~\cite{Fuerst:2018}. 

Li atoms can be paired into Li$_2$ dimers on the repulsive side of the Feshbach resonance.
These molecules are formed by a weak admixture of the highly excited vibrational bound state $X^1\Sigma^+_g(\nu = 38)$, with $\nu$ the vibrational quantum number \cite{Strecker:2003}.  Their binding energy $E_\mathrm{b}={\hbar^2}/({m_\mathrm{Li} a^2})$  depends on the scattering length $a= a_\mathrm{bg} (1+\frac{\Delta B}{B-B_0})(1+\alpha (B-B_0))$ and thus on the magnetic field strength $B$, with $m_\mathrm{Li}$ mass, $\hbar$ Planck's reduced constant, $B_0 = 834.15\,$G, $a_\mathrm{bg}=-1405~a_0$, $\Delta B=300\,$G, and $\alpha=4\times10^{-5}\,$G$^{-1}$, with Bohr radius $a_0$~\cite{Bartenstein:2005}.

For the atom-atom interactions, we use a Lennard-Jones potential
\begin{equation}
	V_{\mathrm{Li}_2}(r_\text{aa}) = -\frac{C_6^\text{aa}}{r_\text{aa}^6}+\frac{C_{12}^\text{aa}}{r_\text{aa}^{12}}, \quad r_\mathrm{aa} = \left|\vec{r}_{\mathrm{a}_1}-\vec{r}_{\mathrm{a}_2}\right|,
	\label{eq:3}
\end{equation}
where $\vec{r}_\mathrm{a_{1,2}}$ are the atom positions and $C_6^\text{aa}$ and $C_{12}^\text{aa}$ the attraction and repulsion coefficients respectively. 

For each scattering event, we  initialize the molecule on a sphere with radius $r_\mathrm{start}=0.5\,\mu$m, large enough to account for potentially large ion-orbits. The ion is initialized in the center of the Paul trap and both ion and molecule velocities are diced from thermal distributions. Also the orientation of the molecule axis is randomized. The Li atoms are initialized in the outer classical turning point  of the molecular potential $r_\text{ct}$ (see Fig.~\ref{fig:mol_size}), where kinetic energy stems from center of mass motion alone. Initially, the molecules do not rotate. The particles are propagated using a  4$^{\mathrm{th}}$ order adaptive Runge-Kutta method until one of the particles leaves a sphere of radius $r_\mathrm{end}\approx1.3\, r_\mathrm{start}$~\cite{Fuerst:2018}.

We identify three scattering channels:
\begin{itemize}
\item[(i)] Molecular ion formation: Li$_2$ + Yb$^+ \rightarrow$ Li + LiYb$^+$

\item[(ii)]  Dissociation: Li$_2$ + Yb$^+ \rightarrow$ Li + Li + Yb$^+$

\item[(iii)]  Quenching: Li$_2$($\nu$) + Yb$^+ \rightarrow$ Li$_2$($\nu$') + Yb$^+$
\end{itemize}
The reaction products are discriminated by calculating the energy of the possible sub-systems Li$_2$ and LiYb$^+$ at the end of the each simulation. The probability for one of the scattering channels $\chi$ is obtained from Monte Carlo sampling of the starting conditions:
\begin{equation}
	P_\chi  =  \frac{N_\chi}{N}\pm \delta_{N_\chi}, \quad 	\delta_{N_\chi} = \sqrt{\frac{N_\chi (N-N_\chi)}{N^3}},
\end{equation}
where $N_\chi$ and $\delta_{N_\chi}$ denote the number and standard deviation of the trajectories associated with channel $\chi$  and $N$ the total number of trajectories. 

The reaction rate is
\begin{equation}
 	\Gamma'_\chi = \kappa \sqrt{T_{\mathrm{Li}_2}}\, (P_\chi \pm\delta_{N_\chi} ),
 	\label{eq:5}
\end{equation}
with $T_\mathrm{Li_2}$  the molecule temperature and $\kappa \approx 3.29\times 10^{-11}$\,m$^3$s$^{-1}$K$^{-\frac{1}{2}}$~ [Appendix~A]. 
For each parameter setting we propagate $2\,\times\,10^4$ trajectories and use temperature distributions around $T_\mathrm{Li_2}=2\,\mu$K and $T_\mathrm{Yb^+}=100\,\mu$K ($\bar{E}_\text{col}/k_\mathrm{B}\approx11\,\mu$K), if not stated otherwise. The average collision energy of the system is calculated with
\begin{equation}\label{eq_Ecoll}
	{\bar{E}}_\mathrm{col} = \frac{\mu}{m_{\mathrm{Yb}^+}}\frac{\mathrm{5}}{2}k_\mathrm{B}T_{\mathrm{Yb}^+} + \frac{\mu}{m_{\mathrm{Li}_2}} \frac{3}{2}k_\mathrm{B}T_{\mathrm{Li}_2},
\end{equation}
where $\mu$ is the ion-molecule reduced mass, and $k_\text{B}$ is the Boltzmann constant. Note, that we account for the intrinsic micromotion of the ion in the Paul trap by counting five degrees of freedom~\cite{Berkeland:1998} in Eq.~(\ref{eq_Ecoll}).

\begin{figure}
	\includegraphics[width=1\columnwidth]{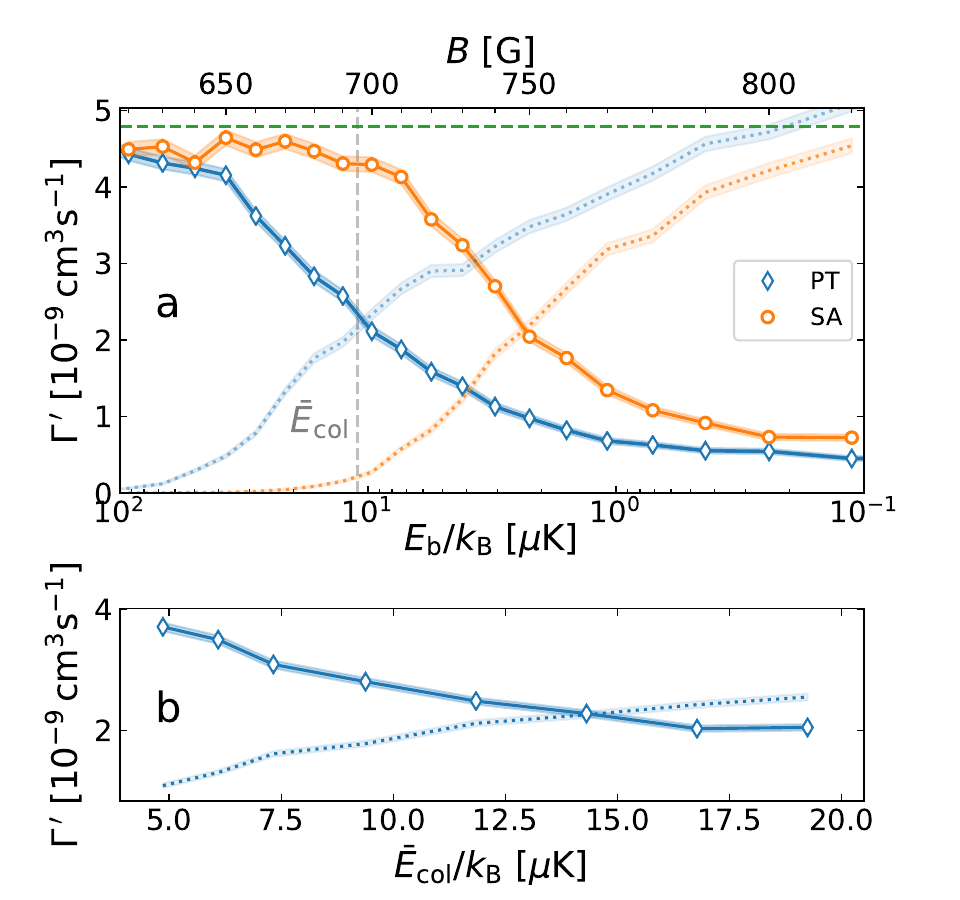}
	\caption{{Reaction rates for molecular ion formation (markers, solid lines) and dissociation (dotted lines). a) Binding energy dependence  for  $\bar{E}_\text{col}/k_\mathrm{B}\approx11\,\mu$K (gray dashed line), corresponding to $T_\mathrm{Li_2}=2\,\mu$K and $T_\mathrm{Yb^+}=100\,\mu$K. In (blue) Paul trap (PT) and in (orange) the time independent secular approximation (SA). Green dashed line is the Langevin collision rate for Li-Yb$^+$. b) Collision energy dependence with fixed binding energy ${E}_\mathrm{b}/k_\mathrm{B}=8.6\,\mu$K for $T_\mathrm{Li_2}=2\,\mu$K and  $T_\mathrm{Yb^+}$ =12.5--100$\,\mu$K in the Paul trap.}}
	\label{fig:eCol}
\end{figure}

\begin{figure}
	\centering
	\includegraphics[width=1\columnwidth]{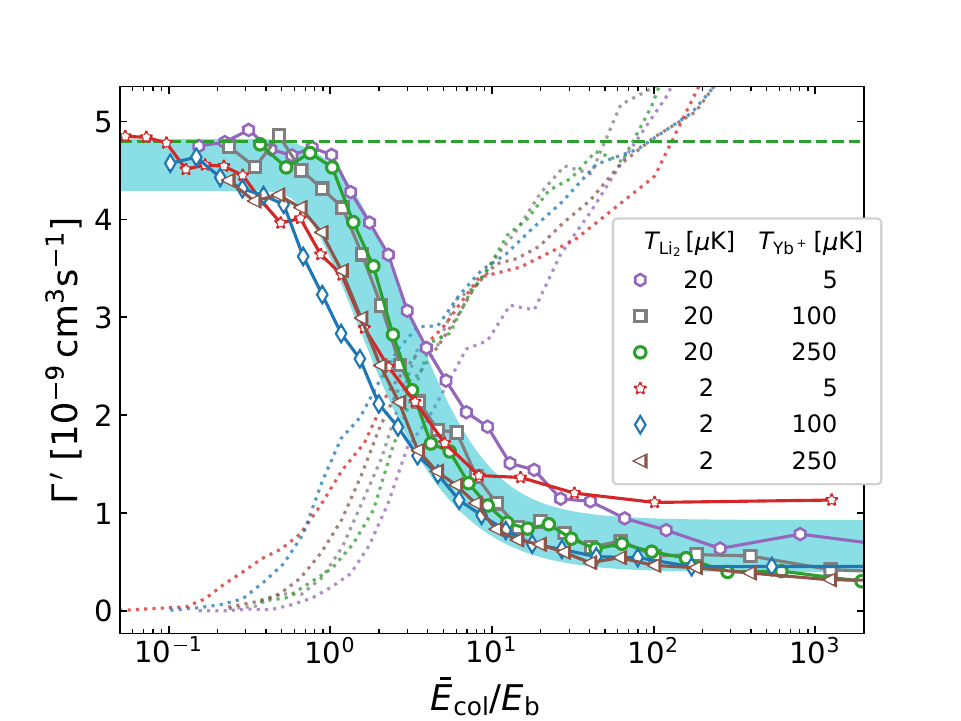}
	\caption{{Reaction rate dependence on the ratio of average collision energy to binding energy in the Paul trap. Molecular ion formation LiYb$^+$ (markers) and dissociation rates (dotted lines) for six different collision energies. The broad blue line is the analytic model described in the text and the horizontal green dashed line indicates the Langevin rate. Error bars ($\lesssim0.3\times 10^{-9}\,\mathrm{cm^3 s^{-1}}$) are omitted for visibility.}}
	\label{fig:eColeBind}
\end{figure}

\section{Results}
We investigate the reaction rates as a function of the Li$_2$ binding energy $E_\mathrm{b}$,
as shown in Fig.~\ref{fig:eCol}a. We compare the reaction rates for molecular ion formation $\Gamma'_{\mathrm{LiYb}^+}$  and  Li$_2$ dissociation $\Gamma'_\mathrm{diss}$ in the Paul trap (PT) and in a time-independent harmonic trap with the same trap frequencies $\omega_\perp$ and $\omega_3$
corresponding to the so-called secular approximation (SA). We find two different regimes. For tightly bound molecules ($E_\mathrm{b}\gg\bar{E}_\text{col}$) molecular ion formation is the dominant channel, with $\Gamma'_\mathrm{LiYb^+}$ close to the Langevin collision rate $\Gamma'_{\mathrm{L}} \approx 4.8\times10^{-9}\,$cm$^3$s$^{-1}$, while $\Gamma'_\mathrm{diss}$ is negligible. Approaching the Feshbach resonance ($E_\mathrm{b}\ll \bar{E}_\text{col}$), dissociation becomes the dominant process, while  $\Gamma'_{\mathrm{LiYb}^+}$ decreases to a roughly constant rate below $1.5\times10^{-9}\,$cm$^3$s$^{-1}$. 
The rate for breaking up Li$_2$-dimers is approximately constant $\Gamma'_\mathrm{diss}+\Gamma'_{\mathrm{LiYb}^+}\approx\Gamma'_\mathrm{L}$ for $E_\mathrm{b}/k_\mathrm{B}\gtrsim 1\,\mu$K. Here, we find that $<5\%$ of the Li$_2$-dimers do not break up in a  Langevin collision~[Appendix A]. 
Close to the Feshbach resonance we find $\Gamma'_\mathrm{diss}>\Gamma'_\mathrm{L}$, as even glancing collisions can dissociate the very weakly bound molecules~[Appendix A]. 
{The location of the crossover strongly depends on the collision energy $\bar{E}_\mathrm{col}$ as can be seen in Fig.~\ref{fig:eCol}b. }
With a fixed binding energy $\bar{E}_\mathrm{b}/k_\mathrm{B}=8.6\,\mu$K and $T_\mathrm{mol}=2\,\mu$K we vary the $T_\mathrm{Yb^+}$ between $12.5-100\,\mu$K.  For small collision energies we find $\Gamma'_\mathrm{diss}<\Gamma'_{\mathrm{LiYb}^+}$. With increasing $\bar{E}_\mathrm{col}$ more tightly bound molecules can be dissociated. Hence, the crossover occurs at larger binding energies.

We investigate the reaction rate dependence on the ratio of collision to binding energy for various ion and atom temperature distributions. 
As shown in Fig.~\ref{fig:eColeBind}, all simulations can be roughly explained by the ratio $\mathcal{\xi}=\bar{E}_\mathrm{col}/E_\mathrm{b}$. For $\mathcal{\xi}\ll1$ the system is in the molecular-ion formation regime. At $\mathcal{\xi}\approx 1$ the dissociation channel opens and $\Gamma'_\mathrm{LiYb^+}$ decreases until dissociation becomes the dominant process for $\mathcal{\xi}\gg1$.
{

To explain this behavior we develop a simple model.
The reaction rate for product $\chi$ can be separated into events from Langevin (L) and non-Langevin (non-L) collisions $\Gamma'_\chi = \Gamma'_\mathrm{L,\chi} + \Gamma'_\mathrm{non-L,\chi}$. However, to form LiYb$^+$ the minimum atom-ion distance must become small and therefore only Langevin collisions can contribute.
The reaction rate is thus given by 
\begin{equation}
 	\Gamma'_\mathrm{L,LiYb^+}\left(\mathcal{\xi} \right) = \Gamma'_\mathrm{L}-\Gamma'_\mathrm{L,Li_2}  - \Gamma'_\mathrm{L, diss}\left(\mathcal{\xi}\right).
\end{equation}
{The dissociation channel 
 opens for $\bar{E}_\mathrm{col} > c\times E_\mathrm{b}$. Here, the factor $c$ takes into account the unknown interchange of the reaction channels as well as the possibility of energy transfer from the oscillating field of the Paul trap.} 
We therefore obtain the ratio of LiYb$^+$ events 
 by integrating the collision energies, which we assume to be Maxwell-Boltzmann distributed, up to $c\times E_\mathrm{b}$
\begin{align}
F_\mathrm{LiYb^+}({ \xi})&=2\int_{0}^{{c\times E_b}} \sqrt{\frac{E'_\mathrm{col}}{{\bar{E}_\mathrm{col}}^3\pi}} \times\exp{\left(-\frac{E'_\mathrm{col}}{{\bar{E}_\mathrm{col}}}\right)} dE'_\mathrm{col}\nonumber\\
	&=  -2e^{-c/\mathcal{\xi}}\sqrt{\frac{c}{\pi \mathcal{\xi}}}+\mathrm{erf}\left(\sqrt{c/\mathcal{\xi}}\right).
	\label{eq:fit}
\end{align}
The reaction rate is then $\Gamma'_\mathrm{LiYb^+}(\xi) = a+b\times F_\mathrm{LiYb^+}(\xi)$, where $a$ is the molecular ion formation rate for $\bar{E}_\mathrm{col}\gg E_\mathrm{b}$ and $b = \Gamma'_\mathrm{L}-\Gamma'_\mathrm{L,Li_2} - a$ relates the model to the Langevin collision rate.  

{Fitting our model to the numerical results, }
{we find good agreement, as can be seen in Fig. \ref{fig:eColeBind} and we extract $a\approx0.7\times10^{-9}\,\mathrm{cm^3s^{-1}}$, $b\approx3.9\times10^{-9}\,\mathrm{cm^3s^{-1}}$ and $c\approx3.1$. We obtain $\Gamma'_\mathrm{L}-\Gamma'_\mathrm{L,Li_2}\approx 4.6\times10^{-9}\,\mathrm{cm^3s^{-1}}\approx0.95\times\Gamma'_\mathrm{L}$, indicating that in only very few Langevin collisions the Li$_2$-dimers do not break up.} 
	}

\begin{figure}
	\centering
	\includegraphics[width=\columnwidth]{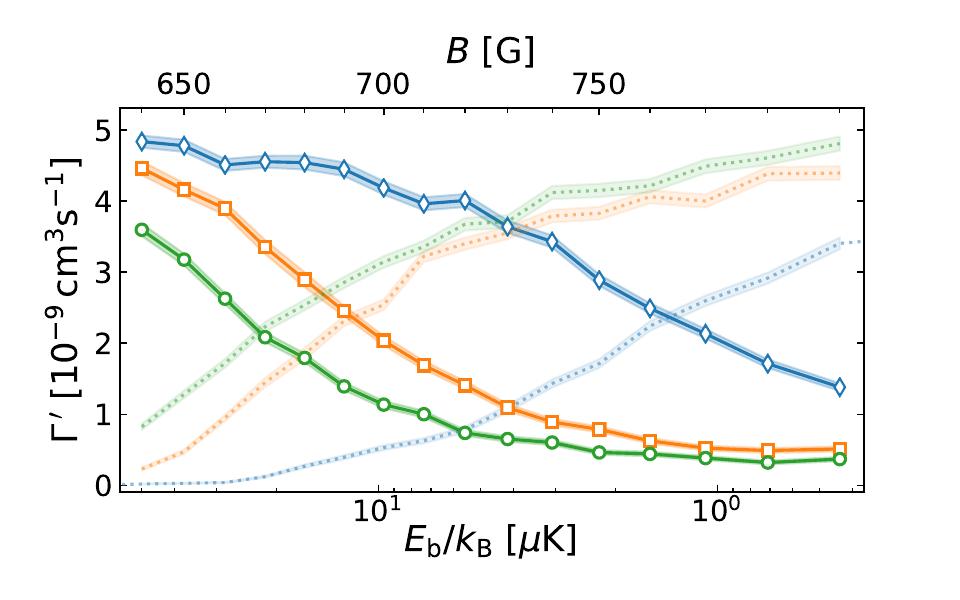}
	\caption{Effect of excess micromotion on the reaction rates for $\bar{E}_\mathrm{col}=11\,\mu$K. Molecular ion formation (markers, solid lines) and Li$_2$ dissociation (dotted lines) for additional electric fields of (blue) $0\,$Vm$^{-1}$, (orange) $0.3\,$Vm$^{-1}$ and (green) $0.6\,$Vm$^{-1}$.}
	\label{fig:harm_emm}
\end{figure}

\section{Influence of the Paul trap}
As shown in Fig.~\ref{fig:eCol}a, the reaction rates show a significant difference between the PT and the SA. In the PT the dissociation channel opens up at larger binding energies compared to the SA. The time-dependent fields change the dynamics of the molecule-ion collisions, such that the PT has the effect of an increased collision energy, compared to the SA. 

Moreover, the PT can suffer from imperfections that cause {so-called excess micromotion (emm) which leads to significantly higher ion kinetic energies. In an experiment, a common cause of emm are stray electric fields that push the ion away from the trap center such that it experiences a non-zero oscillating field. In Fig.~ \ref{fig:harm_emm} we present reaction rates for a PT with additional stray electric fields up to $0.6$\,Vm$^{-1}$ in transverse direction.
We see that introducing emm has the same effect as scattering with a higher collision energy. In state-of-the-art Paul traps, stray electric fields can be eliminated down to $\sim$~0.1~Vm$^{-1}$~\cite{Harter:2013,Keller:2015,Feldker:2020}. 
We conclude that the crossover regime to molecular ion formation should remain observable at realistic emm levels. At the same time, applying well controlled electric fields will allow us to tune the collision energy without influencing the molecule density or the trap depth of the Paul trap as has been demonstrated for atom-ion collisions \cite{Harter:2013,Feldker:2020}.}

\section{Kinetic and binding energy of LiYb$^+$}

\begin{figure}
	\centering
	\includegraphics[width=\columnwidth]{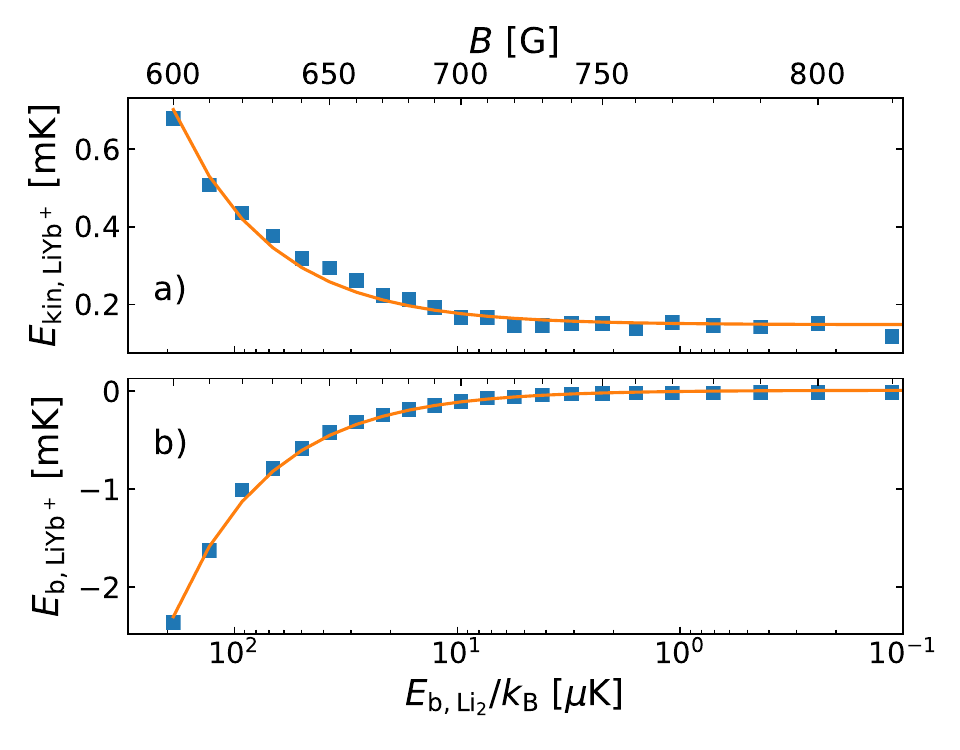}
	\caption{Properties of the created LiYb$^+$ as function of Li$_2$ binding energy. a) LiYb$^+$ (center-of-mass) kinetic energy and b) LiYb$^+$ binding energy. Blue squares show medians to number of occurrences in the numerical simulation and orange lines are linear fits.}
	\label{fig:kinetic_binding_energy}
\end{figure}
We study the (center-of-mass) kinetic energy and binding energy of the created molecular ions for the simulations presented in Fig 2 a) (PT). Therefore, we extract the median of the created LiYb$^+$ energies, see Appendix~B for more details. Fig.~\ref{fig:kinetic_binding_energy} a) shows the kinetic energies of the created LiYb$^+$ as a function of the initial Li$_2$ binding energy. The numerical data is described well by the linear correlation
\begin{equation}
	E_\mathrm{kin,LiYb^+} \approx 0.1\,\mathrm{mK} +  2.9\times E_\mathrm{b, Li_2}.
\end{equation}
With energies below $1\,$mK the presented method indicates its potential for creating ultracold molecular ions. 

In Fig.~\ref{fig:kinetic_binding_energy} b) the resulting LiYb$^+$ binding energies are shown as a function of the initial Li$_2$ binding energy. The molecular ions are weakly bound with binding energies in the mK regime. We find the linear correlation
\begin{equation}
	E_\mathrm{b,LiYb^+} \approx 0.01\,\mathrm{mK} - 12.3\times E_\mathrm{b, Li_2}.
\end{equation}

\section{Discussion \& Conclusion}
Our simulations reveal the existence of a crossover regime from Li$_2$ dissociation to LiYb$^+$ formation in collisions of a trapped ion with Feshbach dimers. {Importantly, the full crossover is within experimental reach as the required collision energies and magnetic field strengths have been reported~\cite{Feldker:2020}.} We find that the created molecular ions have kinetic energies with a median below $1\,$mK and thus are easily trapped for typical Paul trap depths $\gg 10$\,K. The created LiYb$^+$ are weakly bound with binding energies on the order of mK. From comparing to BaRb$^+$, a radiative lifetime on the order of $2\,$ms for a binding energy of about $1\,$mK can be expected~\cite{Mohammadi:2020}.


It is appealing to study both theoretically and experimentally how quantum effects will appear as deviations from our model. In particular, the crossover to the quantum regime for atom-ion collisions occurs for $\bar{E}_\mathrm{col}\approx E^*=\hbar^2/(2\mu_\text{ai} (R^*)^2)$, with $R^*=\sqrt{2\mu_\text{ai} C_4^\text{ai}/\hbar^2}$. For $^{171}$Yb$^+$-$^6$Li we find $E^*/k_\text{B}=8.6$~$\mu$K, which becomes equal to the molecular binding energy for $B\approx705\,$G.  
A significant increase in richness can be expected once the collision energy of the atoms and ion reach deep in the quantum regime, where Feshbach resonances between these particles can play a role as well~\cite{Tomza:2015}.

Our results point to a novel method for creating ultracold molecular ions. 
It will be interesting to study buffer gas cooling of the formed molecular ion with the  atomic gas or ultracold collisions of molecular ions with Feshbach dimers. {The LiYb$^+$ molecular ion has a large permanent electric dipole~\cite{Tomza:2015,DaSilva:2015}} and may allow the study of dipole-dipole interactions. By cotrapping an atomic Yb$^+$ ion, the molecular ion can be straightforwardly identified by mass spectrometry using, e.g., the collective modes of ion motion. These techniques can also be used to perform quantum logic spectroscopy and to study the properties of the molecular ion \cite{Wolf:2016,Chou:2017, Sinhal:2020}.

\section*{Acknowledgements}
We would like to acknowledge Thomas Feldker for inspiring this work. We gratefully thank Henning F{\"u}rst, Micha\l\ Tomza and Jook Walraven for fruitful discussions. This work was supported by the Netherlands Organization for Scientific Research (Vidi Grant 680-47-538 and Start-up grant 740.018.008 (R.G.), and Vrije  Programma 680.92.18.05) (G.C.G., R.G., J.P.). R.S.L. acknowledges funding from the European Union’s Horizon 2020 research and innovation
programme under the Marie Sklodowska-Curie grant agreement No 895473.

\section*{Appendix A: Model checks}
\label{AppA}

We present additional information for {controlling the nature of a charged impurity in a bath of Feshbach dimers.} The parameters used for the numerical simulations are presented in Tab.~\ref{tab:parameters}, if not stated differently. For the Paul trap the values correspond to the experimental ones from Ref.~\cite{Feldker:2020} and the atom-ion coefficients are taken from Ref.~\cite{Fuerst:2018}. For the Feshbach molecules we used Ref.~\cite{Bartenstein:2005}, with this, the magnetic fields deviate $<1\%$ from recent measurements~\cite{Zuern:2013}. First, we explain how the reaction rates are obtained from individual collisions and we compare the Langevin collision rate from our numerical simulation with the analytically expected Langevin rate. We look at Langevin and non-Langevin collisions and study their reaction rates separately. The Li$_2$ model potential is discussed and a convergence check is presented for the chosen potential depth. Finally, we do an energy conservation check and investigate the tolerance of our $4^\mathrm{th}$ order Runge-Kutta stepper method.

\begin{table}
	\centering
	\begin{tabular}{lcr}\\
		\hline
		Coefficient 			& description 		& Value\\
		\hline
		$f_\mathrm{z}$		& axial trap frequency 	& $130.037\,$kHz 	\\	
		$f_\mathrm{rf}$		& rf-drive frequency	& $1.85\,$MHz	\\
		$q_{x,y}$		& rad. $q$-parameter	& $\pm0.5$\\
		$q_z$		& ax. $q$-parameter	& 0 	\\
		$T_\mathrm{ion}$		& initial ion temp.	& $5-250\,\mu$K	\\
		$T_\mathrm{mol}$	& mol. temp.	& $2-20\,\mu$K	\\
		$r_\mathrm{start}$				& mol. start radius	& $0.5\,\mu$m \\
		$r_\mathrm{end}$				& mol. escape radius	&	$0.655\,\mu$m\\
		$p_\mathrm{tol}$		& rel. num. tolerance	& 	$10^{-10}$\\ 
		$C_4^{\mathrm{ai}}$			& attractive 	&	$5.607\times10^{-57}\,$J m$^4$	\\
		$C_6^{\mathrm{ai}}$			& repulsive 	&	$5\times 10^{-19}\,$m$^2\times C_4$    \\
		$C_6^{\mathrm{mol}}$		& attractive 		& 	$1.3\times 10^{-76}\,$J m$^6$ \\
		$C_{12}^\mathrm{mol} $	& repulsive		& 	$3.2\times 10^{-128}\,$J m$^{12}$\\
		$B$						& magn.-field strength		&  	$600-830\,$G\\
		$E_\mathrm{b}/k_\text{B}$		& mol. binding energy & $0.003-187\,\mu$K\\
		\hline
		
	\end{tabular}
	\caption{Parameters used in the simulations if not stated differently.}
	\label{tab:parameters}

\end{table}

\begin{figure}
	\includegraphics[width=\columnwidth]{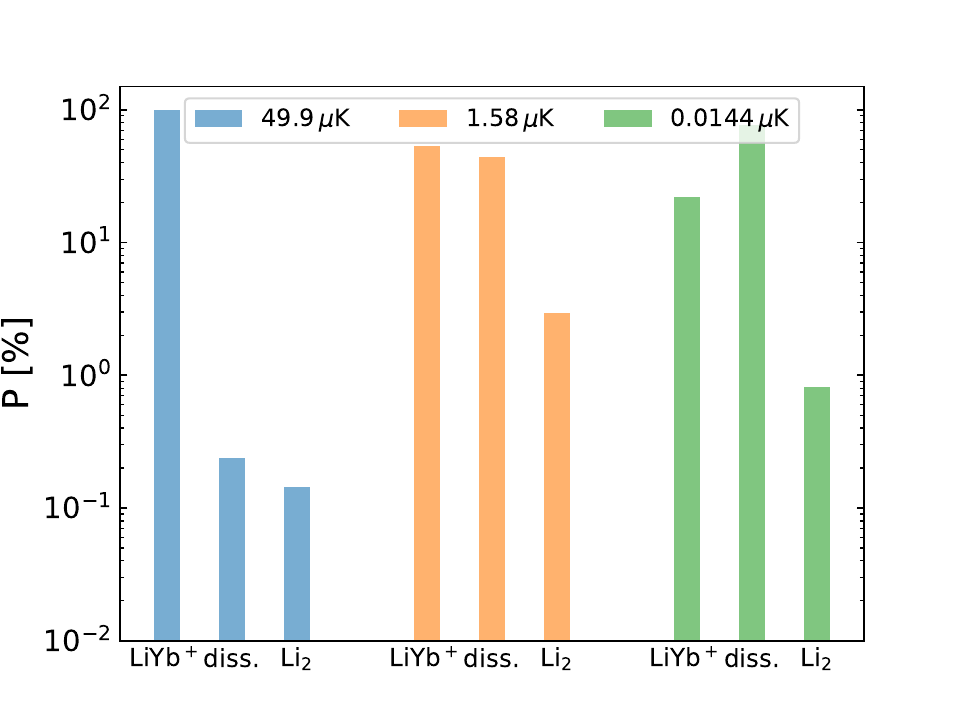}
	\caption{Percentage of reaction rates from Langevin collisions for $E_\mathrm{b}/k_\mathrm{B} = 49.9\,\mu$K, $1.58\,\mu$K and $0.0144\,\mu$K.}
	\label{fig:langevin_reaction_rate}
\end{figure}

\begin{figure}
	\includegraphics[width=\columnwidth]{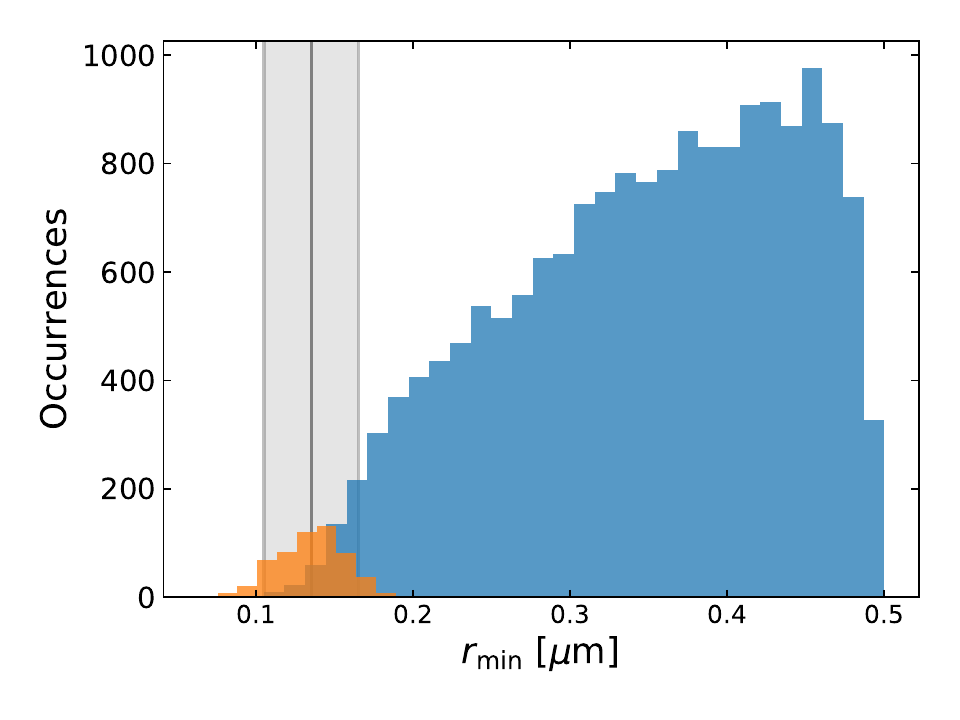}
	\caption{Occurrences of minimal atom-ion distance for non-Langevin collisions at $E_\mathrm{b}/k_\mathrm{B}=0.0144\,\mu$K. Reactions with Li$_2$ maintained (blue) and reactions with Li$_2$ dissociated (orange). Break up occurs only close to Langevin impact parameter (gray line), where the linewidth of the shaded gray area indicates two times the molecule size.}
	\label{fig:histogram}
\end{figure}

\subsection*{1. Extracting  rates from the numerical simulations}
In the numerical model we simulate single ion-molecule collisions and count the different reaction products. To extract a rate from these results, we do a projection onto an atomic (molecular) density by using
\begin{equation}
	V_\mathrm{s}\frac{\braket{v}_\mathrm{s}}{\braket{s}_\mathrm{s}}=\kappa\sqrt{T_{\mathrm{Li}_2}},
\end{equation}
where the average velocity from a Maxwell-Boltzmann distribution is given by
\begin{equation}
	\braket{v}_\mathrm{s} = \sqrt{\frac{8 k_\mathrm{B} T_{\mathrm{Li_2}}}{\pi m_{\mathrm{Li}_2}}},
\end{equation}	
with molecule temperature $T_\mathrm{Li_2}$ and Boltzmann constant $k_\mathrm{B}$.
For a sphere of volume $V_\mathrm{s}=\frac{4}{3} \pi r_\mathrm{start}^3$ with radius $r_\mathrm{start}$ the average distance for two points on the surface is $\braket{s}_\mathrm{s} = \frac{4}{3} r_\mathrm{start}$. We obtain
\begin{equation}
	\kappa =  \sqrt{\frac{8 \pi k_\mathrm{B}}{ m_{\mathrm{Li}_2}}} r_\mathrm{start}^2.
	\label{eq:kappa}
\end{equation}
For $r_\mathrm{start}=0.5\,\mu$m used in all simulations presented in the main text we obtain $\kappa\approx 3.29\times 10^{-5}\mathrm{cm}^3\mathrm{s}^{-1}\mathrm{K}^{-\frac{1}{2}}$. Then, the reaction rates are obtained with Eq.~5 of the main text.

For a consistency check, we compare the Langevin collision rate from our simulation with one obtained analytically. We  simulate about $1.2\times10^6$ atom-ion collisions with $r_\mathrm{start}=0.3\,\mu$m. We monitor the minimal atom-ion distance $r_\mathrm{min}$ during the entire propagation and label a collision Langevin if $r_\mathrm{min}<5\times10^{-9}$\,m. Note, that due to the spiraling character of the Langevin collisions the exact value for the discrimination has no significant influence. We find about $0.24\times 10^6$ Langevin collision. Using  Eq. \ref{eq:kappa} (and Eq.~5 from the main text), we extract the numerical Langevin (L,n) rate of
\begin{equation}
	\Gamma'_\mathrm{L,n}=4.71\pm0.01\times 10^{-9}\,\mathrm{cm^3s^{-1}}.
\end{equation}
The analytical (a) Langevin collision rate is calculated from
\begin{equation}
	\Gamma'_\mathrm{L,a} = 2 \pi \sqrt{\frac{C_4}{\mu}}\approx  4.80\times10^{-9}\, \mathrm{cm}^3\mathrm{s}^{-1}.
\end{equation}
The two rates agree within $\sim2\%$, showing that the simulations are consistent with the analytical model.

\begin{figure}
	\includegraphics[width=\columnwidth]{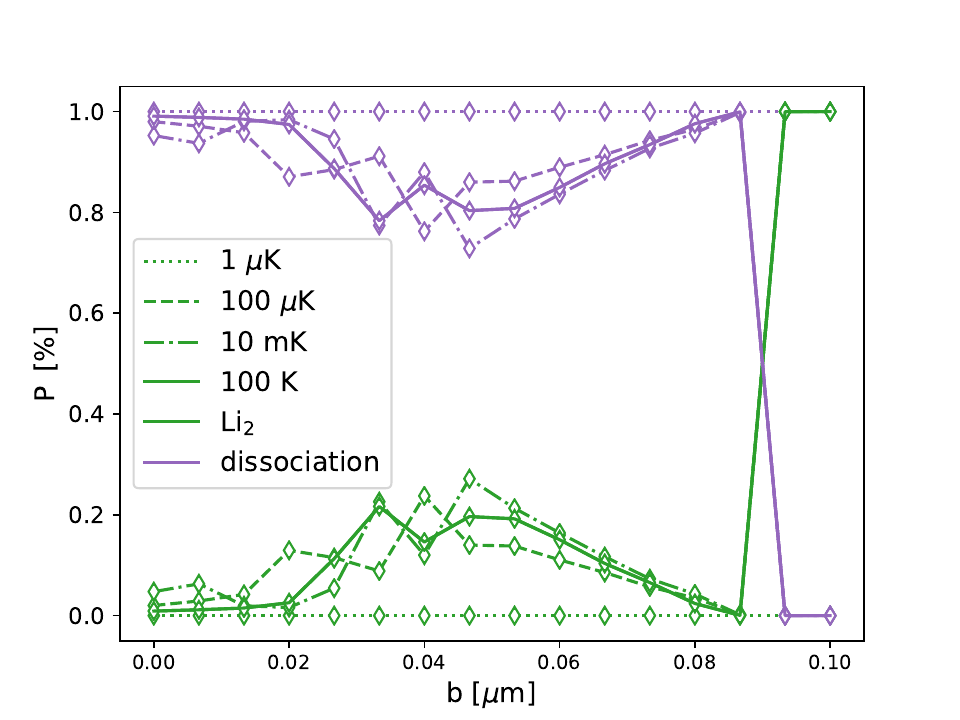}
	\caption{Reaction probability dependence on impact parameter $b$ and without Paul trap for different depths of the Li$_2$ potential $E_\text{pd}/k_\text{B}= {1\,\mu \mathrm{K}-100\,\mathrm{K}}$. The molecule (purple lines) gets dissociated or (green lines) survives the collision.}
	\label{fig:pot_depth}
\end{figure}

\subsection*{2. Langevin and non-Langevin collisions}
In the main text we find for large binding energies, that the reaction rates for breaking up a Li$_2$-dimer are similar to the Langevin collision rate. However, close to the Feshbach resonance, the dissociation rate even exceeds the Langevin rate, as can be seen in Fig.~2a (and Fig.~3) of the main text. Here, we separate Langevin from non-Langevin collisions to study their reaction rates independently. We label a collision as Langevin if the minimal atom-ion distances  $r_\mathrm{min}<5\times10^{-9}$\,m, for at least one of the two atoms. We run $2\times10^4$ collisions with ($r_\mathrm{start}=0.5\,\mu$m) for $49.9$, $1.58$ and $0.0144\,\mu$K from which roughly $10\%$ are Langevin collisions. For the Langevin collisions the reaction rates are  shown in Fig. \ref{fig:langevin_reaction_rate}. The probability of an unbroken Li-dimer from a Langevin collision is roughly $P_\mathrm{L,Li_2}= (0.14, 2.95, 0.81)\,\%$ for $(49.9, 1.58, 0.0144)\,\mu$K. 

For non-Langevin collision (not shown) we find no dissociation events for $49.9$ and $1.58\,\mu$K. For weaker bound molecules at $0.0144\,\mu$K we find $P_\mathrm{non-L,diss.} \approx3\%$. Since about $90\%$ of the simulation are non-Langevin collisions, this results in a significant amount, such that $\Gamma'_\mathrm{diss}>\Gamma'_\mathrm{L}$ for weakly bound dimers. 

We find that dissociation in non-Langevin collisions only occurs close to the edge of the critical impact parameter $b_\text{c}=(2C_4^\text{ai}/E_\mathrm{col})^{1/4}$ (see main text). This is shown in Fig.~\ref{fig:histogram}. There, the occurrences of $r_\mathrm{min}$ for dissociation (orange) and unbroken Li-dimers (blue) are shown for non-Langevin collisions at $0.0144\,\mu$K. The collisions leading to dissociation have small $r_\mathrm{min}$ around the Langevin impact parameter (center gray line), indicating that glancing collisions contribute to the dissociation rate low binding energies.

\begin{figure}
	\includegraphics[width=\columnwidth]{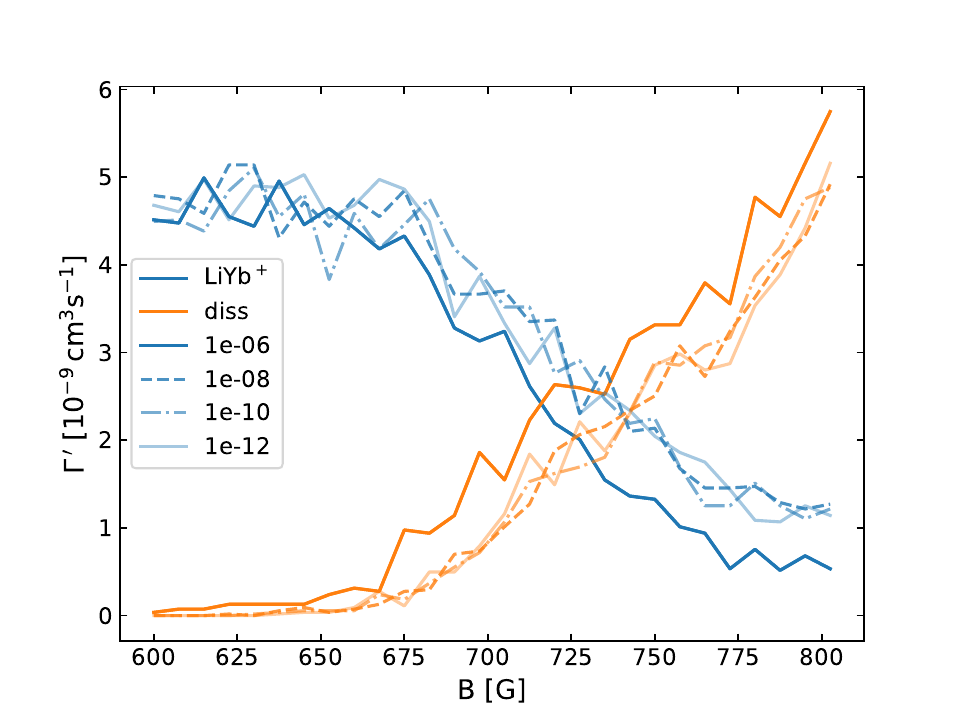}
	\caption{Reaction rates for various relative numerical tolerances of the stepper method with $T_\mathrm{Yb^+}=5\,\mu$K and $T_\mathrm{Li_2}=20\,\mu$K.
}
	\label{fig:tolerances}
\end{figure}
\begin{figure}
	\centering
	\includegraphics[width=\columnwidth]{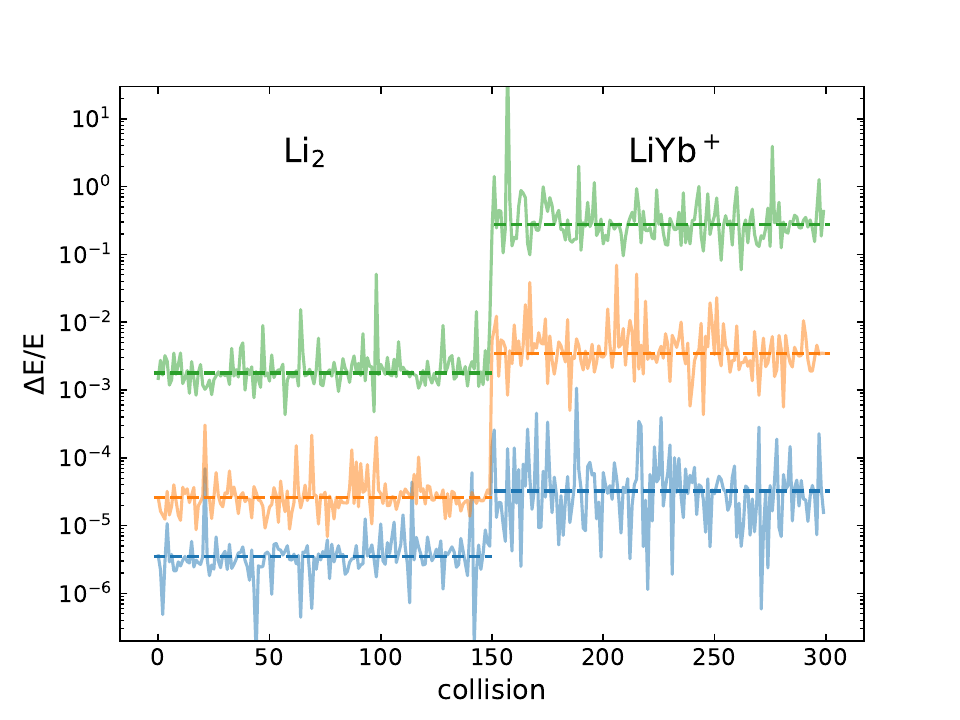}
	\caption{Energy conservation for simulations without trapping fields for different tolerances $10^{-12}$ (blue), $10^{-10}$ (orange) and $10^{-8}$ (green). Collision 0-149 result in Li$_2$ and 150-300 in LiYb$^+$. The step in accuracy likely stems from the large difference of energy scales in the LiYb$^+$ system.}
	\label{fig:energy_conservation}
\end{figure}

\subsection*{3. Molecule model potential}
\label{depth}

The model potential (Eq.~3 main text) for the molecules was chosen to be rather shallow and we present here a consistency check to verify that this does not influence our results. The shallowness of the potential is important to reduce the required computational effort: The ion moves on length scales of $\lesssim0.5\,\mu$m ($T_\mathrm{ion}=250\,\mu$K), and thus a reasonably large simulation sphere has to be chosen. To avoid the propagation of very quickly-oscillating molecules over these large distances, it is beneficial to use a very shallow potential for the molecule.  This is possible since the physical behavior should be dominated by the long-range term. Here, we choose the $C_{12}^\text{aa}$ coefficent of Eq.~3 in the main text to limit the potential depth to $E_\mathrm{pd}/k_\text{B} = 10\,$~mK. 
We do a convergence check without the Paul trap and with $T_\mathrm{ion}/k_\text{B}=0$\,K. The binding energy is $E_\mathrm{b}/k_\text{B}\approx 0.1\,\mu$K. We launch the molecules from a fixed starting position and only randomize different molecule orientations towards the ion. The result is  shown in Fig.~\ref{fig:pot_depth}. We see no significant deviations in a range from $100\,\mu$K to $100\,$K on the shown reaction rates.

\subsection*{4. Tolerance of the stepper method}

We perform a convergence test to verify the numerical tolerance used  for the Runge-Kutta stepper method. Therefore, we do a full simulation for different tolerances, see Fig.~\ref{fig:tolerances}. The reaction rates are independent of the tolerance up to $10^{-8}$. Only for a tolerance of  $10^{-6}$ do we see the rates deviate, especially for higher $B$-fields. We chose $10^{-10}$ for all simulations.

\subsection*{5. Conservation of energy and angular momentum}
\begin{figure}
	\centering
	\includegraphics[width=\columnwidth]{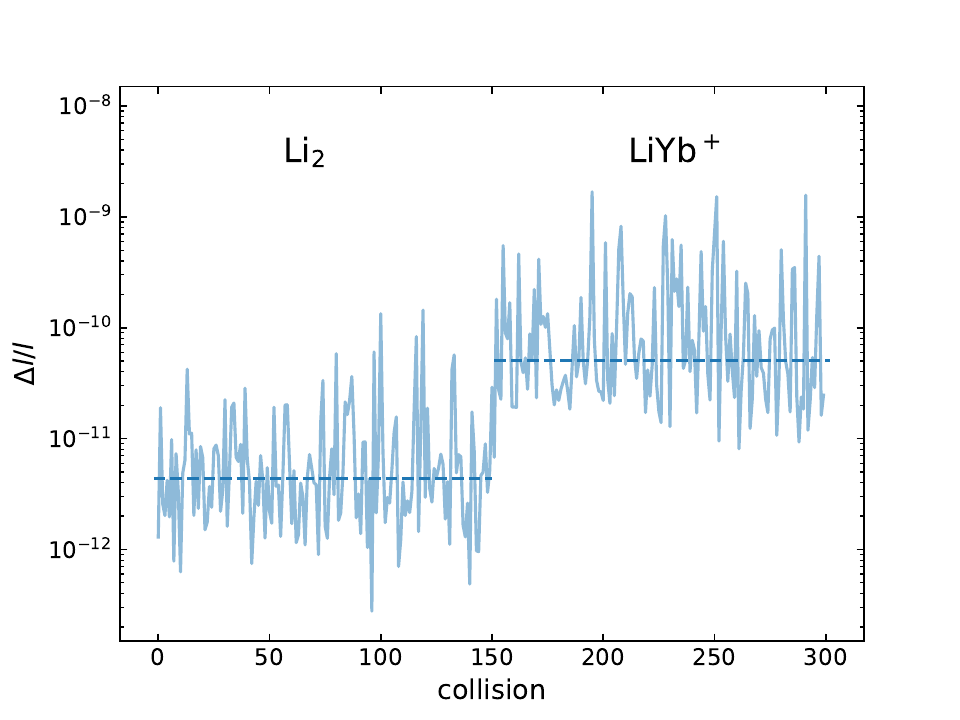}
	\caption{Angular momentum conservation for simulations without trapping fields and for a tolerance of $10^{-10}$. Collision 0-149 result in Li$_2$ and 150-300 in LiYb$^+$. }
	\label{fig:angular_momentum_conservation}
\end{figure}

Finally, we check the conservation of energy and angular momentum. Since time dependent fields can pump energy into the system during a collision~\cite{Cetina:2012}, we simulate collisions without the ion trap.

The relative change in energy $\Delta E/E$ is shown in Fig.~\ref{fig:energy_conservation} for different tolerances $10^{-8} - 10^{-12}$ of the stepper \cite{Fuerst:2018}. For a tolerance of $10^{-10}$ we find $\Delta E/E\lesssim10^{-4}$ for collisions that do not break up the Li$_2$-dimer, while collisions leading to LiYb$^+$ have a lower accuracy. This we attribute to the stiffness of the LiYb$^+$ systems due to the different time scales of the fast motion of the bound Li and the slow propagation in the trap. For lower tolerances the accuracy increases by a factor of roughly 10, at the cost of approximately doubling the computation time. 

Similarly, we compute the change in total angular momentum $\Delta l/l$ during the collisions, with $l = |\vec{l}|=\sum_i{\vec{r_i}\times\vec{p_i}}$ and the sum over the three particles. The result is shown in Fig.~\ref{fig:angular_momentum_conservation} for a tolerance of $10^{-10}$. We find $\Delta l/l< 10^{-8}$ for all recorded events. As in the energy conservation test we see a step in accuracy for the different reaction channels.\\

\section*{Appendix B: LiYb$^+$ binding energy}

\label{AppB}
\begin{figure}[h]
	\centering
	\includegraphics[width=\columnwidth]{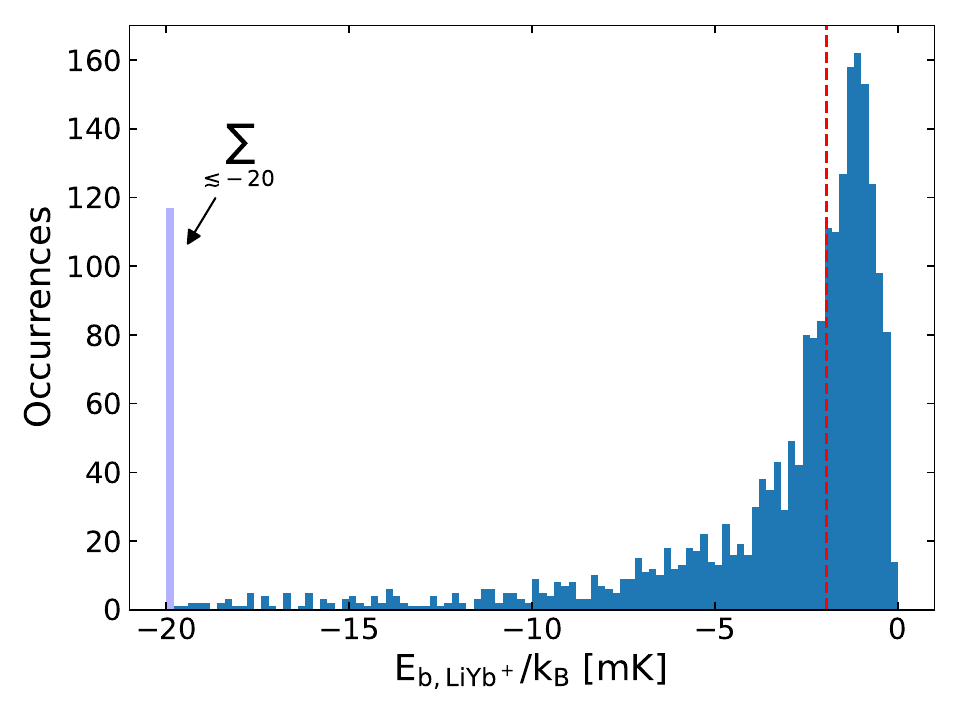}
	\caption{LiYb$^+$ binding energies for T$_\mathrm{Yb+}=100\,\mu$K, T$_\mathrm{Li_2}=2\,\mu$K and $E_\mathrm{b}/k_\mathrm{B} = 187\mu$K. The red dashed line shows the median for the observed occurrences.}
	\label{fig:binding_energy}
\end{figure}

We look into the binding energies of the created LiYb$^+$ for the system on the molecular ion creation side of the crossover. Therefore, we use T$_\mathrm{Yb+}=100\,\mu$K, T$_\mathrm{Li_2}=2\,\mu$K and $E_\mathrm{b}/k_\mathrm{B} = 187\mu$K. The occurrences of the observed energies are summarized in Fig~\ref{fig:binding_energy}. We find, that for these settings approximately $89\%$ of LiYb$^+$ have binding energies above $-10\,$mK and we extract a median (red dashed line) of about $-1.95\,$mK.



%

\end{document}